# The Beer Can Theory of Creativity

Liane Gabora
University of British Columbia

**OUTLINE**

## 1. INTRODUCTION

I had to laugh this morning while reading the *Ottawa Citizen* when I saw a crafty yet bafflingly incompetent vandal described as "He's got a full six pack, but the plastic thingy that holds them together is missing". It's a spin-off of the saying "He's one can short of a full six pack", which itself is a Canadianized version of "He's lost a few marbles" or "He's not playing with a full deck". Not only does the newspaper description beautifully exemplify one of the main issues of this chapter-the interplay of variation and continuity as a creative insight is adapted from one context or circumstance to another. But content-wise, it's a pithy summary of another, related issue dealt with here: in order to *adapt* the idea to a new context, in order to *evolve* it in new directions, it must originally have been stored in memory in a way that implicitly identifies its relationships to *other* ideas. In other words, when it comes to creativity, how your 'beer cans' are connected together is as important as how many of them there are.

This chapter explores the cognitive mechanisms underlying the emergence and evolution of cultural novelty. Section Two summarizes the rationale for viewing the process by which the fruits of the mind take shape as they spread from one individual to another as a form of *evolution*, and briefly discusses a computer model of this process. Section Three presents theoretical and empirical evidence that the sudden proliferation of human culture approximately two million years ago began with the capacity for creativity that is, the ability to generate novelty strategically and contextually. The next two sections take a closer look at the creative process. Section Four examines the mechanisms underlying the fluid, associative thought that constitutes the inspirational component of creativity. Section Five explores how that initial flicker of inspiration crystallizes into a solid, workable idea as it gets mulled over in light of the various constraints and affordances of the world into which it will be born. Finally, Section Six wraps things up with a few speculative thoughts about the overall unfolding of this evolutionary process.

## 2. CULTURE AS AN EVOLUTIONARY PROCESS

When asked in an interview (LeMay 1990) "What is creativity?" Russian dancer Rudolf Nureyev gave the following answer:

> It is something born from within you. It's as though you felt a need to do something, to say something, to utter, and you cannot live without uttering this sentence or writing this piece of music. It just begs to manifest itself. It is a need to express yourself first, and then to rationalize this expression. It is irrational first, rational after. I am sure Einstein had an inkling about something unknown and then came to his theory of light. And I am sure everybody has had this impulse, very much akin to sex, sexual drive, or sexual appetite, if you wish.

It is fascinating that even nonscientists frequently describe their creative process through metaphorical allusion to how novelty arises in the biological world. Is this merely tactical? For example, since everyone has experienced sexual drive, does comparing the creative impulse to it lull us into a romanticized view of the extent to which the creative impulse possesses one who gives their life over to it, not voluntarily but because that is their nature? And when artists use the word 'birth' to describe the process by which something vaguely intuited, and perhaps seemingly irrational, is nurtured into the rational realm of human events and understandings, are they manipulating us into viewing their works with the reverence with which we view a newborn child? Or is there really a deep isomorphism between these two novelty-generating processes?

### 2.1 Variation and Convergence in Biology and Culture

Increasingly, culture is being viewed as a form of evolution (e.g. Campbell 1974, 1987; Cavalli-Sforza & Feldman 1981; Csikzentmihalyi 1993, 1999; Gabora 1996, 1997; Popper 1963). To see why, we need to take a closer look at what evolution is: a process wherein a stream of information incrementally adapts to environmental constraints. It requires:

- A *pattern of information* (an entity that occupies a state within a space of possible states).
- A means of *varying* the pattern (exploring or transforming the space).
- A rationale for selecting or *converging* on variations that are adaptive, i.e. that tend to give better performance than their predecessors in the context of some problem or set of constraints (a fitness landscape applied to the space).

With each iteration of variation and convergence, the evolving entity is better able to meet the challenges and capitalize on the opportunities presented by its circumstances; in a sense it becomes a sharper 'mirror image' of its environment. The situation is complicated by the fact that the entity's environment often consists largely of other entities which are *themselves* evolving, so symbiotic alliances and coevolutionary arms races become established.

In biological evolution, the evolving patterns of information are genes encoded as sequences of nucleotides. *Variation* arises through mutation and recombination, and natural selection weeds out those that are maladaptive. In cultural evolution, the evolving patterns of information are concepts, ideas, attitudes, values, *etc*. Variation is generated by creatively combining, transforming, and restructuring them. Factors promoting convergence include biological drives, goals, desires, values, aesthetic preferences, and the associative organization of memory, which constrains how one concept evokes (in a sense, selects) another. Actually, in the cultural domain, variation and convergence tend to go hand-in-hand; this process will be explored in detail shortly.

The cultural analog of the *genotype* is the network of concepts *etc*. that together constitute a model of reality, or worldview. The analog of *phenotype* is the way they get implemented or communicated, through facial expressions, gestures, actions or vocalizations. Implementation incorporates syntactic features characteristic of the channel through which it is conveyed (Brooks 1986); thus, for example, a dance step looks different with each individual who performs it. Whereas biological information gets passed on to offspring as a complete set of genes, cultural information spreads one idea at a time.

The assimilation of new ideas into a society alters the selective pressures and constraints it exerts on the individuals embedded in it, which in turn alters the generation and proliferation of future ideas. Thus culture, like biology, comprises a self-sustained system for the exploration and transformation of a space of possible patterns.

## 2.2 Is More than One Mind Necessary for Ideas to Evolve?

Sometimes a fourth item is included in the list of requirements for evolution: a way of replicating (or amplifying, as molecular biologists refer to it) the selected variations. Ideas are said to replicate through social processes such as teaching and imitation. Clearly replication is an integral component of biological evolution, and the transmission of ideas through social processes such as imitation and teaching is likewise important. But is it *indispensable?* If, for example, you were the only human left on the planet, but you could live forever, would cultural evolution grind to a halt? If you found an ingenious way to scale a mountain, you would still have come up with something new, something more adapted to the environment, something which you might go on to modify and perfect¾ to evolve. Your novel mountain-scaling approach might even lead you, through metaphorical analogy, to new ideas about how to elevate your mood, or expose you to mountain plants and thereby lead to novel ways of cooking them, weaving them, *etc*. Thus, strictly speaking, there need not necessarily be more than one individual for culture to evolve. Evolution can occur through variation and convergence acting on a single stream of information (such as a train of thought) without anything being explicitly replicated. There is no more reason to refer to this as replication than to refer to your cat *now* as a replicant of the cat that jumped on the window sill a few minutes ago. The conscious experience of an individual can be viewed as an evolutionary process in which variation and convergence are not spatiotemporally separated but intimately intertwined, one pattern of qualia fluidly transmuting into the next.

Nevertheless, the culture of a single individual would be impoverished, to say the least. Cultural variety increases exponentially as a function of the number of creative, interacting individuals. As a simple example, a single individual who invents ten new words is stuck with just those ten. A society of ten interacting individuals, only one of whom is creative, is no better off; there are still just ten words. In a society of ten nonsocial individuals, each of whom invents ten words but does not share them, each individual still has only ten words. But in a society where the ten individuals invent ten words and share them all, everyone ends up with a hundred words.

The bottom line is: culture as we know it, with its explosive array of meaningful gestures, languages, and artifacts, requires the kind of parallel processing that social interaction provides. But technically, a single individual can evolve cultural novelty.

## 2.3 Meme and Variations: A Computer Model of Cultural Evolution

This can be seen clearly in Meme and Variations (MAV), a computer model of cultural evolution (Gabora 1995). The program consists of an artificial society of interacting neural network based agents which do not have genomes, and neither die nor have offspring, but which invent, implement, and imitate ideas, or memes. Every iteration, each agent has the opportunity to acquire a new meme, either through 1) *innovation*, by mutating a previously-learned meme, or 2) *imitation*, by copying a meme implemented by a neighbor. Whereas memes do not evolve at all in the absence of innovation, meme evolution *does* take place in the absence of imitation (albeit more slowly). In other words, when agents can generate novelty, but completely ignore one another, memes still evolve.

MAV displays many phenomena found in biological evolution, such as drift, and slower increase in fitness at epistatic loci. It also displays phenomena unique to culture. For example, *mental simulation* (ability to assess the relative fitness of a meme before actually implementing it) and *strategic innovation* (using past experience to bias how memes mutate, as opposed to mutating at random) both increase the rate at which fitter memes evolve.

The higher the ratio of innovation to imitation, the greater the meme diversity, and the higher the fitness of the fittest meme. However, meme fitness increases most rapidly for the society as a whole with an innovation to imitation ratio of 2:1 (but diversity is then compromised). Interestingly, for the agent with the fittest memes, the less it imitates (i.e. the more computational effort reserved for its own creative efforts), the better it performs.

The approach taken in MAV can be compared with Sims' (1991) and Todd and Latham's (1992) computer models of the evolution of creativity which, though they explore a *cultural* process, use a genetic algorithm−−a model of *biological* evolution. However, although MAV is modeled after cultural evolution, it is too simple to explore many cultural phenomena. The space of possible memes is fixed and small, and (unlike real life) the fitness function that determines what constitutes a good meme is predetermined and never changes. Also, imitation and innovation are probably not as discrete in real life as MAV would suggest. Nevertheless, it demonstrates the feasibility of computationally modeling the processes by which creative ideas spread through a society giving rise to observable patterns of cultural diversity.

### 2.4 Breadth-first versus Depth-first Exploration

We have seen in the preceding sections, 2.2 and 2.3, that it is not essential for individuals to imitate (or that there even be more than one of them!) for culture to evolve. The following distinction can be useful for gaining insight into why this is so.

Biological creativity is largely random. Billions of mutant gametes are ejected into the world, and most are unsuccessful, but the occasional one is better than the average, and over generations it tends to increase in the population. This kind of approach−−search the entire space of possibilities without devoting much effort to any one possibility, and probably at least one of them will be better than what exists now−−is referred to in computer science as *breadth-first.*

Human creativity, on the other hand, is highly non-random. Say, for example, you are faced with the problem of how to stop a leaky faucet from dripping. If you were to try to reach a solution by mutating each feature of the dripping faucet one by one, or by recombining it with every concept from hotdogs to postmodern deconstruction, you wouldn't get very far. We generate novelty *strategically*, using an internal model of the relationships amongst the various elements of the problem domain, and *contextually*, responding to the specifics of how the present situation differs from previously encountered ones. This kind of approach−−explore few possibilities, but choose them wisely and explore them well−−is referred to as *depth-first.*

In a sense, culture embodies the best of both worlds; that is, each individual's depth-first stream of thought is embedded in a highly parallel, relatively breadth-first social matrix which provides a second, outer tier of convergent pressure. For depth-first search to be successful, of course, requires some knowledge of the topology of the fitness landscape; in other words, some smarts about what sort of variation and convergence would probably be beneficial. Of course, even a relatively breadth-first algorithm like biological evolution operates with a certain degree of smarts. In fact it is sometimes argued that it too is highly nonrandom, though clearly it is not strategic and contextual in the way a stream of thought is. So the distinction is just a matter or degree.

**2.5 Dampening Arbitrary Associations and Forging Meaningful Ones**

Whether or not an algorithm is breadth first or depth first, the notions of linkage equilibrium and disequilibrium are useful conceptual devices for gaining an overview of what is taking place during an evolutionary process. The closer together two genes are on a chromosome, the greater the degree to which they are *linked*. *Linkage equilibrium* is defined as random association amongst alleles of linked genes. Consider the following simple example:

*A* and *a* are equally common alleles of Gene **1**.

*B* and *b* are equally common alleles of Gene **2**.

Genes **1** and **2** are linked (nearby on same chromosome).

There are four possible combinations of genes **1** and **2**: *AB*, *Ab*, *aB*, and *ab*. If these occur with equal frequency, the system is in a state of linkage equilibrium. If not, it is in a state of linkage *dis*equilibrium. Disequilibrium starts out high, but tends to decrease over time because mutation and recombination break down arbitrary associations between pairs of linked alleles. However, at loci where this *does not* happen, one can infer that some combinations are fitter, or more adapted to the constraints of the environment, than others. Thus when disequilibrium does not go away, it reflects some structure, regularity, or pattern in the world.

What does this have to do with creativity? Like genes, features of memories and concepts are connected through arbitrary associations as well as meaningful ones. We often have difficulty applying an idea or problem-solving technique to situations other than the one in which it was originally encountered, and conversely, exposure to one problem-solving technique interferes with the ability to solve a problem using another technique (Luchins 1942). This phenomenon, referred to as mental set, plays a role in cultural evolution analogous to that of linkage in biological evolution. To incorporate more subtlety into the way we carve up reality, we must first melt away arbitrary linkages amongst the discernable features of memories and concepts, thereby increasing the degree of equilibrium. This needn't be a particularly intellectual process; for example, the *feeling* of a particularly painful or joyous experience could be extricated from the specifics of that experience, and re-manifest itself as, say, a piece of music. As we destroy patterns of association that exist because of the historical contingencies of a particular domain, we pave the way for the forging of associations that reflect genuine structure in the world of human experience which may manifest in several or perhaps all domains.

Nureyev revealed an at least intuitive grasp of this when, further on in the interview discussed near the beginning of this section, he was asked "Has going from (dance to acting) given you any insight into the previous kind of performance styles, and into the different shades of creativity"? He responded: "If you know one subject very well, then you have the key to every other subject." If this is true, it doesn't seem particularly fair. Why should some people hold, not just the key to the discipline they have specialized in, but *all* the keys? However, looking around, there appears to be increasing evidence in support of Nureyev's claim. Complexity theory is being applied to everything from earthquake prediction, to neuroscience, to the design of an algorithm for animating the behavior of a flock of bats in Batman. Genetic algorithms are being used to compose music, and mathematicians are turning into fractal artists. It isn't obvious what doors will be opened by any particular key!

## 2.6 Order, Chaos, Interconnectedness, and Information

Before moving on, there is one more important point to be made about evolutionary systems in general, and that is the following. The degree to which the parts of an information-evolving system are correlated or causally connected falls squarely in a narrow regime between order and chaos (Langton, 1992). Another way of expressing this is to say that the mutual information--that is, the amount of information that can be gleaned about one component of a system by examining another component--must be intermediate. To use an analogy that conveys why, think of a sequence of identical, uniformly-spaced drum beats. Each beat provides information, but because an entire sequence of beats can be recoded as the single, albeit slightly more complex instruction "loop drum", it isn't very informative. On the scale from complete order to complete chaos, it lies at the extreme order end. On the other hand, if each instrument in a band plays notes randomly without regard for what preceded or what will follow, or for what the other instruments are doing, this can be recoded as 'play anything'. Thus, at the extreme chaos end, the system is also informationally stagnant. Evolutionary systems are poised at the proverbial edge of chaos where the degree of connectivity is intermediate.

Let us consider how this principle manifests in biological systems. Organic polymers such as protein and RNA molecules commonly act as ingredients or catalysts of some chemical reactions, and products of others. The more reactions catalyzed per polymer, the more interconnected the system. If each enzyme catalyzed only one reaction, the specificity of the system would be so high that once it worked there would be no room for improvement. However, if each enzyme catalyzed each reaction equally well, the system would be completely unstable. The actual situation is that, much as different keys sometimes open the same door but require different amounts of effort, reactions can be catalyzed by many catalysts, with varying degrees of efficiency. Kauffman (1993) found that living systems, and boolean network models of them, attain the delicate edge of chaos state of intermediate connectedness without top-down control through spontaneous self-organization.

Let us now consider the implications of the edge of chaos principle for cognition. If memories were distributed so narrowly (that is, stored in such tiny portions of the mind) that their storage regions never overlapped, the current experience would have to be *identical* to a previously-stored one to evoke it. However, if they were *too* widely distributed, successive thoughts would not necessarily be meaningfully related to one another. The free-association of a schizophrenic seems to correspond to what one might expect of a system like this (Weisberg, 1986). In fact, when patterns are distributed *throughout* a neural network or holograph, unless they are perfectly orthogonal they interfere with one another, a phenomenon known as crosstalk. For the mind to be capable of evolving a stream of meaningfully related yet potentially creative remindings, the degree of distribution must be between these extremes; that is, the sphere of concepts activated by any stimulus must fall within an intermediate range. Thus, a given sensory input activates not just *one* location in memory, nor does it activate *every* memory location to an equal degree, but activation is *distributed* across many memory locations, with degree of activation falling with distance from the most activated one.

This feature is sometimes incorporated into neural networks using a radial basis function (RBF) (Hancock et al., 1991; Holden & Niranjan, 1997; Lu et al. 1997; Willshaw & Dayan, 1990). Each input activates a hypersphere of memory locations, such that activation is maximal at the center *k* of the RBF and tapers off in all directions according to a (usually) Gaussian distribution of width s , as in Figure 1. The further a stored concept is from *k*, the less activation it not only *receives* from the stimulus input but in turn *contributes* to the stimulus output, and the more likely its contribution is cancelled out by that of other simultaneously evoked locations. A wide s in effect models the situation where neurons have a lower activation threshold, so more fire in response to a given stimulus. In neural networks, suitable values for *k* and s are found during a training phase. In the brain the requisite tuning of patterns of neuron interconnectivity is probably achieved through self-organizing feedback processes (Edelman, 1987; Pribram 1994).

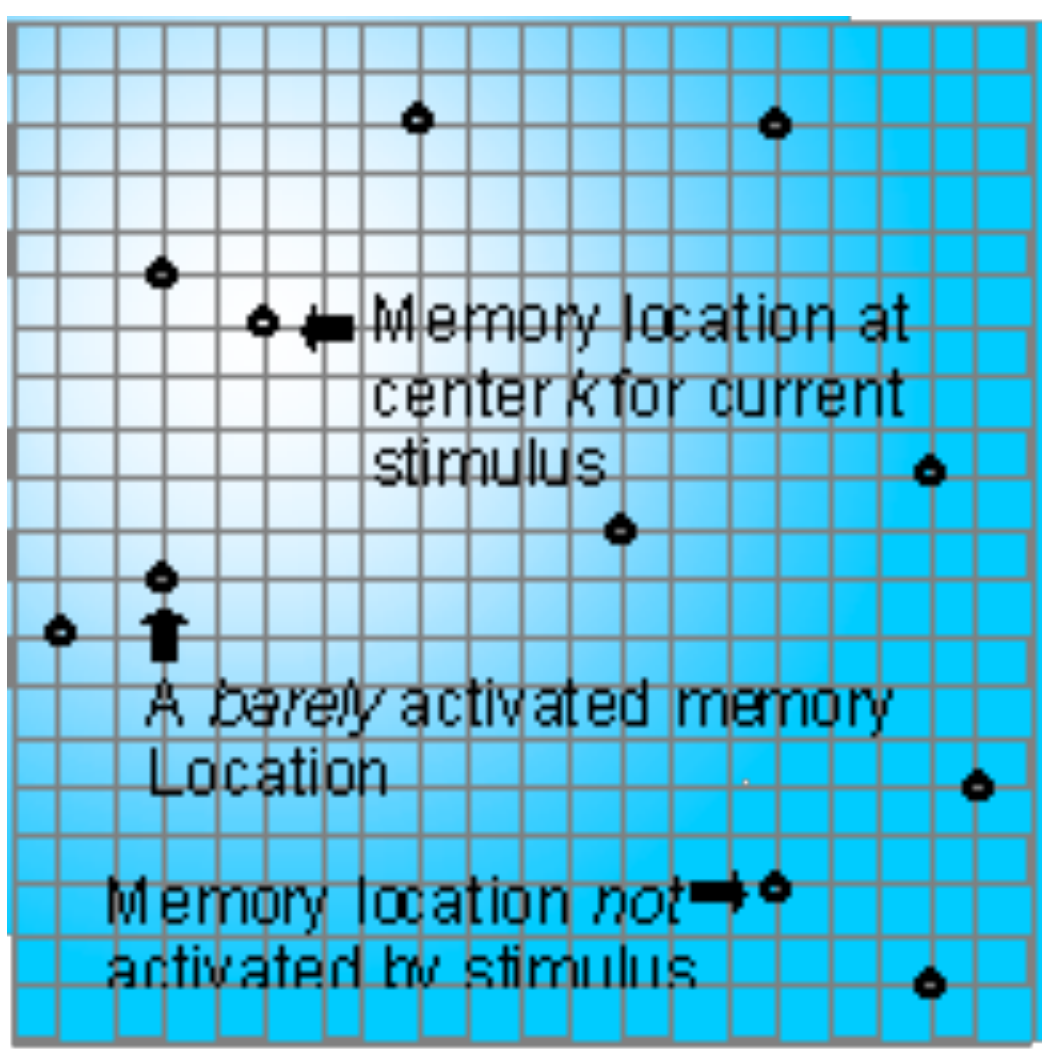

Figure 1. Highly schematized diagram of a stimulus input activating two dimensions of a memory or conceptual space. Each vertex represents a possible memory location, and black dots represent actual location in memory. Activation is maximal at the center *k* of the RBF, and tapers off in all directions according to a Gaussian distribution of width s .

The result of achieving the delicate edge-of-chaos state is that sequential slices of a stream of experience are *self-similar*, that is, correlated but non-identical; each is a *variation* of its predecessor. Self-similarity is enhanced by the fact that the motivational state generally does not change unless the present goal is satisfied, or another becomes more pressing, and also by the high degree of continuity in our percepts (for instance, if a window is to your left *now* it is likely to still be there *now*).

To briefly sum up thus far, we have examined the rationale behind viewing culture as a form of evolution: a process by which information incrementally adapts to environmental constraint through variation and convergence. In some ways these two forms of evolution operate quite *differently;* for example, cultural novelty is generated in a more depth-first manner. In other respects they are similar, and features of biological evolution−−such as drift, linkage equilibrium, and the edge of chaos principle−−transfer readily to culture. We now turn from the general principles underlying these evolutionary systems to the question of how they began in the first place.

## 3. CREATIVITY AS THE ORIGIN OF CULTURE

The origin of culture is sometimes unquestioningly equated with onset of the capacity for social learning, and particularly imitation (e.g. Blackmore 1999). But the idea that imitation is what makes us human seems counterintuitive. When we feel proud to belong to the human race, we think of the great pyramids, beautiful music, the airplane: in short, the fruits of *creativity*. The word 'imitation' is, in fact, often used to denote inferiority.

Are our intuitions about what makes us special actually misguided? I think not. There are several sources of evidence that creativity, not imitation, is what makes us human.

### 3.1 Theoretical Evidence

We saw in the MAV computer model that when the agents' ability to imitate is set to maximum, and their ability to invent is turned off, nothing happens. This makes sense. There has to be something worth imitating before the ability to imitate comes in handy, or even manifests itself. A society of individuals who can imitate, but not invent, is stagnant. Therefore, the suggestion that it was the appearance of imitation that brought about the onset of culture is not theoretically sound.

On the other hand, recall how when the MAV agents' ability to *invent* is set to maximum, and their ability to *imitate* is turned off, evolution does take place, albeit more slowly than with imitation. Since, as we have seen, cultural evolution is more depth-first than biological evolution, once the capacity for creativity presents itself, culture can evolve, whether or not there is imitation. Novelty can then breed more novelty. Or as one choreographer (whose name I forget) put it: "If we don't do what our predecessors did, we're doing what our predecessors did." Thus the proposal that culture originated with the onset of creativity is at least theoretically possible.

### 3.2 Archeological Evidence

Human culture is generally thought to have originated somewhere around 1.7 million years ago, during the time of *Homo erectus*. This period marks the appearance of sophisticated stone tools and habitats, use of fire, long-distance hunting strategies, and migration out of Africa, as well as a rapid increase in brain size (Bickerton 1990; Chase 1990; Chase & Dibble 1987, 1992; Corballis 1991; Donald 1991). This increase in variety of artifacts and habitats is exactly what one would expect to see if humans suddenly acquired the capacity to be creative. It is the *opposite* of what one would expect if they suddenly acquired the ability to imitate. If it were imitative capacity that had suddenly arisen, then just prior to this time there would have been a great variety of tools, habitats, *etc*., and the onset of imitation would have funneled this variation in just a few of the most useful directions. Thus the archeological evidence is consistent with the thesis that creativity, not imitation, was the bottleneck to culture.

### 3.3 Evidence from Animal Behavior

Experimental research indicates that imitation (and related phenomena such as emulation, response facilitation, *etc*.) are widespread in the animal kingdom. It has been documented in budgerigars (Galef et al. 1986), quail (Akins & Zentall 1998), cowbirds (King & West 1989), rats (Heyes & Dawson 1990; Heyes et al. 1992), monkeys (Beck 1976; Hauser 1988; Nishida 1986; Westergaard 1988), orangutans (Russon & Galdikas 1993) and chimpanzees (Goodall 1986; Mignault 1985; Sumita et al. 1985; Terrace, Petitto, Sanders & Bever 1979; Waal 1982; Whiten 1998). Nevertheless, as many authors have pointed out, although imitation is commonplace, no other species has anything remotely approaching the complexity of human culture (e.g. Darwin 1871; Plotkin 1988).

As we have seen, imitative capacity remains latent, hidden from view, until there is variation for it to work on. Thus the lack of cultural complexity in animals despite evidence that, when put to the test, they can imitate, is consistent with the "creativity as bottleneck to culture" proposal.

### 3.4 What Caused the Onset of Creativity?

What then could have caused humans to suddenly be capable of generating strategic, contextual novelty? For a stream of creative thought to unfold, related memories and sensorimotor behaviors (some hardwired, some learned) must become woven into an interconnected conceptual web, or worldview. However this presents the following paradox. Until a mind incorporates *relationships* between memories, how can one thought evoke another? And until one thought can evoke another, how are relationships established amongst memories so that they become an interconnected worldview?

The origin of life presents an analogous paradox: if living organisms come into existence when other living things give birth to them, how did the first organism arise? That is, how did something able to reproduce itself come to be? By combining an insight from random graph theory (Erdos & Renyi 1959, 1960) with the concept of hypercycles (Eigen & Schuster 1977, 1978, 1979), Kauffman (1993) arrived at the hypothesis that life may have begun with, not a single molecule capable of replicating *itself*, but an *autocatalyticaly closed* set of *collectively* self-replicating molecules. (Note that it is *not* closed in the sense that new molecules cannot be incorporated into the set. It is closed in a mathematical sense, but not a physical sense.)

An analogous line of reasoning can be applied to explain how discrete memories become woven into a worldview. Although this account focuses on integration of the worldview through the emergence of deeper, more general concepts, the principles apply equally to integration of the psyche through the purification of intentions and emotions. A detailed account of the proposal can be found in (Gabora 1998), and elaborations in (Gabora 1999, 2000), but the basic line of reasoning goes as follows. Much as catalysis increases the number of different polymers, which in turn increases the frequency of catalysis, reminding events increase concept density by triggering *abstraction*--the formation of abstract concepts or categories such as TREE or BIG--which in turn increases the frequency of remindings. And just as catalytic polymers reach a critical density where some subset of them undergoes a phase transition to a state where there is a catalytic pathway to each polymer present, concepts reach a critical density where some subset of them undergoes a phase transition to a state where each one is retrievable through a pathway of remindings events or associations. Finally, much as autocatalytic closure transforms a set of molecules into an interconnected and unified living system, conceptual closure transforms a set of memories into an interconnected and unified worldview. Memories are now *related* to one another through a network of abstract concepts; the more abstract the concept, the greater the number of other concepts that fall within a given distance of it in conceptual space and therefore are potentially evoked by it. For instance, your concept of DEPTH is deeply woven throughout the matrix of concepts that constitute your worldview; it is latent in experiences as dissimilar as DEEP SWIMMING POOL, DEEP-FRIED ZUCCHINI, and DEEPLY MOVING BOOK.

In Section 2.6 we saw that to produce a stream of meaningfully-related yet potentially creative remindings, memories need to be distributed (though their distribution must be constrained). The process described here would most likely have been kickstarted by a genetic mutation leading to decreased neuron activation threshold, causing the storage and retrieval of memories to become more widely distributed.

So, to return to the Bob and Doug MacKenzie parlance with which this chapter began, culture may have begun with the emergence of a *plastic thingy*--a hierarchical network of abstract concepts that connects associated memories into a worldview. Some experiences are either so consistent, or so inconsistent, with this worldview that they have little impact on it. Others percolate deep, renewing our understanding of myriad other concepts or events. The worldview is stable if it fosters thought trajectories that enhance individual wellbeing. Much as biological organisms assimilate the food necessary for maintenance and growth but shield off toxins, an individual assimilates stimuli that expand its worldview, but censors or represses stimuli or memories that could bring harm.

### 3.5 A Relationally Structured Conceptual Network Aids both Creativity and Imitation

A relationally and hierarchically structured worldview would be invaluable in biasing the generation of novelty in directions that are likely to be fruitful. How this could work will be discussed in depth in the sections that follow. Actually, a relationally and hierarchically structured worldview probably aids imitation too, especially the imitation of actions, vocalizations, or artifacts that are particularly complex (Byrne & Russon 1998). For example, a novel mannerism or expression could be generated by putting your own slant on it on the fly, *as* you imitate. So the dichotomy between creativity and imitation may less polarized than the preceding discussion implies. Nevertheless, imitation *without* creativity is not sufficient to bring about cultural evolution.

## 4. THE COGNITIVE ARCHITECTURE UNDERLYING CONCEPTUAL FLUIDITY

> These precious things let them bleed let them wash away these precious things let them break their hold over me...
>
> −−Tori Amos, from the song *Precious Things* on the album *Under the Pink*.

I decided to open this section with the above excerpt because it seems to both illustrate, and metaphorically refer to, the experience of conceptual fluidity. Taken in the context of the rest of the lyrics, the 'precious things' she speaks of appear to be painful memories, which are precious in the sense that they are guarded. And to "let them bleed" or "wash away" seems to mean to dissipate their emotional valency by dropping this guard, allowing them to surface from memory and participate in the natural flow of fluid thought, if not in their painful original form, then in the form of a song. Of course, this interpretation is purely speculative and may not be what Tori Amos intended consciously or unconsciously. At any rate, let us now examine the conceptual meltdown of sorts that occurs during the initial inspirational component of the creative process.

### 4.1 The Drive to Unite or Reconcile

Behind biological novelty lurks the tension that comes of the desire to unite male with female, and this tension may also underlie much novelty in the cultural realm, such as the writing of love songs. But we have all also experienced the tension that drives efforts to creatively unite or reconcile seemingly disparate elements of our worldview. This happens, for example, when inexplicable phenomena or experimental results disclose gaps in our understanding of the physical world. Repressed material could also be viewed as a fenced-off region of conceptual space that leaves a gap of sorts in the worldview. Gaps can also arise in our understanding of the motives of others, such as in the cognitive dissonance that occurs when others give us mixed messages (as when there is dissonance between the content of what someone says, and their tone of voice, facial expression, gestures, or body language).

Thus, even once we have attained conceptual closure, new paradoxes and inconsistencies inevitably arise to reveal weaknesses in our conceptual network. If our biological structure can impel us to mate, nurture offspring, and generally evolve biological form, then it doesn't seem impossible that conceptual structure impels us to develop, share, and evolve conceptual form, and that the Eureka high is a sort of 'conceptual orgasm'.

### 4.2 What is an Inkling?

Before we get to the question of what is an inkling, or an exceptional Eureka type of experience, let us briefly examine what happens during a more mundane sort of instant of experience. A perceived stimulus or imaginary construct evokes−−due to spatiotemporal patterns in the constellation of activated concepts at that instant−−an impression, observation, interpretation, judgement, emotional reaction, or assessment of some kind. For simplicity any such response can be referred to as a *measurement* (Aerts & Gabora 1999). We can now turn this around and say that the measurement of a stimulus (or imaginary construct) *reveals* the distribution of probabilities inherent in the state of the conceptual network prior to the stimulus (or imaginary construct). The sequence of situations one finds oneself in can thus be viewed as a stream of measurements that *collapse* the conceptual network into specific configurations or states. You could say that, between measurements, the beer tops flip open and beer starts to ooze out, and the wider the distribution function, the greater the probability that any particular drop of beer will collapse into a different can from the one it was in at the time of the previous collapse. Each measurement then brings to light a thought or feeling that was latent in the worldview, forced into existence an experience it was potentially capable of having.

In a sense, the measurement tests the integrity of a certain portion of the worldview, the size of the portion tested depending on the distribution function (and thus on the degree of entanglement). At the risk of mixing metaphors, you could say it's like throwing a ball against a wall and observing how it responds. The more flexible the material the ball is made of, the more it gives when it makes contact. Similarly, the wider the distribution function, the greater the portion of the worldview that makes contact with the world at that instant. Much as irregularities in the bounced ball cause its path to deflect, constrictions (repressed memories) or gaps (inconsistencies) in the 'collapsed' portion of the worldview may cause tension and thereby indicate a need for creative release, revision, or reconstruction. (Of course, so long as the ball doesn't completely deflate and slide down the wall, you're doing fine. :-)

An *inkling*, then, can be viewed as collapse on an association or relationship amongst memories or concepts that, although their distribution regions overlap, were stored there at different times, and have never been simultaneously evoked before in the same collapse.

### 4.3 Evoking from Memory is a Contextual, Reconstructive Process

Thus the interpretation of an instant of experience, be it of the mundane or the Eureka variety, draws upon the relational structure of the memory that constitutes an individual's worldview. However drawing upon memory does not necessarily imply the faithful recollection of a specific remembered episode. There is a saying "You never step into the same stream twice", and this applies to streams of thought as well as streams of water. At a high enough level of resolution, it is not the exact same memory or concept conjured up time and again; your understanding of it is always colored by, and reinterpreted in the context of, events that have taken place since the last time you thought of it, and your current goals or desires. For example, right now I am recalling how, last night as I was working on this, my cat Inkling jumped up and fell asleep on my lap. Tomorrow I may retrieve the "same" memory. But today it is colored by today's mood, today's events; tomorrow it will experienced slightly differently. The qualia patterns are not identical. In fact, what is retrieved may be never have been explicitly stored. Next month I might wrongly remember it as having been my other cat, Glimmer, perhaps because I will blend this memory with a memory of Glimmer walking on my keyboard. The mind does not pull items from memory like mitts from a box, but creatively weaves external stimuli with memories and concepts relevant to the current motivational state. Thus it is more accurate to think of the process as *reconstruction* rather than retrieval**.**

It is worth stressing that although this process is sometimes construed as *search*, it is just information flowing through a system displaced from equilibrium. The current instant of experience activates certain neurons, which in turn activate certain other neurons, which leads to the distributed storage of that experience, which activates whatever *else* is stored in those locations, which then merges with salient information from senses and drives to form the *next* instant of experience*, et cetera*, recursively, in a stream of what Karmiloff-Smith refers to as representational redescription (1992). Since correlated qualia patterns get stored in overlapping locations, what emerges is that the system appears to retrieve experiences that are similar, or concepts that are relevant, to the current experience. This has a unifying effect on the memory that counters the diversifying effect of new sensory input. In other words, when experience *A* evokes a memory of experience *B*, *B* gets tinged by *A* and vice versa; the subjectivities of these events become integrated, and the system becomes more contextually structured.

Another field where contextuality arises, and where it has been extensively analyzed, is quantum mechanics. My colleagues and I are finding that mathematical formalisms and concepts developed for coping with contextuality in quantum structures are useful for developing a formal description and analysis of the creative mind (Aerts et al., 1999, 2000; Aerts & Gabora 1999).’ Recall how the concept of DEPTH was stored in memory locations that stored other concepts ranging from DEEP SWIMMING POOL to DEEPLY MOVING BOOK. Similarly, the concept of CONTAINER does not just *activate* concepts like CUP, BAG, and so forth; it derives its very *existence* from them. Likewise, once CUP has been identified as an instance of CONTAINER, it is forever more affected, however subtly, by experiences that activate CONTAINER. It is in this sense that concepts exhibit the quantum property of *nonlocality*. Another concept from quantum mechanics that can be fruitfully applied to the mind is that of *entanglement*. For the mind to be an entangled system, it is not enough that concepts be fuzzy or overlap. The mind is entangled because the entire network of concepts overlaps through and through, such that the meaning and subjective feeling associated with any one of them lies in its relations to the meanings and feelings associated with the others.

The bottom line is that the creative mind is a contextual system (even displaying phenomena associated with contextuality in the micro-world, the domain where contextuality has been studied most extensively) and it draws upon memories in a reconstructive manner.

### 4.4 The Pre-Inkling Courtship

What is it about the mind that makes it capable of experiencing inklings? This question can be made more explicit with an example. Since we seem to have a beer theme going here, imagine that you are the graphic designer and chief executive of advertising at Goosehead Breweries, and it is your job to come up with a name and label for a new papaya-flavoured beer. You look out your window at the bleak Saskatchewan landscape and see disappointingly little out there to put you in a tropical frame of mind. You taste the beer itself, and feel even less inspired.

The taste of the beer registers as a vector of features that determines which synapses are excited and which are inhibited, which determines how activation flows through your memory network, which in turn determines the multidimensional hypersphere of locations where this beer taste experience is stored. Since it is an experience of *beer*, it activates memory locations containing related concepts like THIRST-QUENCHING, BEVERAGE, LAGER and ALE. Since it is *papaya* beer, it simultaneously activates concepts like FRUIT, PEAR-SHAPED, JUICY, and TROPICAL. The process of etching this experience *to* these locations triggers release *from* them of whatever else has been stored in them. In other words, the locations that store one instant of experience go on to provide the constituents of the next instant of experience. Of course, nothing is drawn from them if, at that instant, some stimulus or biological drive becomes pressing, such as a phone call, or the need to go to the bathroom. But to the extent that memory contributes to the next instant of awareness, storage of the PAPAYA BEER experience activates retrieval of not only the PAPAYA BEER experience itself, but all other memories stored in the same locations, and the next experience can be calculated by determining the contributions of these memories feature-by-feature. The other evoked memories contribute less to the next instant of experience than the PAPAYA BEER experience you just had, not just because they are less recent, but because they do not lie at the center of the hypersphere of activated locations, and so are statistically likely to cancel one another out. In this instance, the various activated concepts do *not* completely cancel out, at least not with respect to their sound. Since the words TROPICAL and ALE sound alike, the AL sound emerges as a significant dimension.

Perhaps a more sophisticated way of describing this event is to invoke a dynamic perspective, wherein perceptual / cognitive / cultural information carries temporal changes in the nested covariations or bundled phase relations of stimuli (Cariani 1995, 1997; Gabora submitted), and different kinds or dimensions of this information are carried by different frequencies (like a radio broadcast system). Under this scheme, the AL sound in TROPICAL constructively interfere and resonate with the AL sound in ALE.

At any rate, whether it happens through activation of features or resonance of phase relations, the emergence of the AL sound as a significant dimension is an artifact, since the AL sound has seemingly nothing to do with PAPAYA BEER. However, since you know that catchy product names sometimes capitalize this sort of artifact, your memory is probed again with an instant of experience that is the same as the previous one except for heightened activation of the AL sound feature (or phase relation). The first probing, the one that collapsed on an interpretation of the PAPAYA BEER taste experience, was not a waste of time; it got you closer in conceptual space to where you needed to be. The second probing is even more successful. It evokes the new construct, TROPICALE.

The point is that it is because TROPICAL and ALE had both, at different times in the past, been etched into overlapping memory locations--those most attuned to the AL sound--that the collapse on TROPICALE was possible. Though it is a reconstructed blend, something you have never actually experienced, it can still be said to have been evoked from memory. It is certainly not a *profound* inkling. But it is an inkling nonetheless, and good enough to let you to hold onto your job at Goosehead Breweries.

### 4.5 Defocused Attention, Flat Associative Hierarchies, and Conceptual Fluidity

Now the question becomes: what sort of mind would have stored the sound of the word ALE within reach of BEER, and the sound of the word TROPICAL within reach of PAPAYA and come up with a brand name that creatively combined them?

Martindale (1999) has identified a cluster of attributes associated with creativity which includes: defocused attention (Dewing & Battye 1971; Dykes & McGhie 1976; Mendelsohn 1976), high sensitivity (Armstrong 1974; Martindale 1977), and flat associative hierarchies (Mednick 1962), and sensitivity to subliminal impressions (Smith & Van de Meer 1994). The steepness of an individual's associative hierarchy is measured experimentally by comparing the number of words that individual generates in response to stimulus words on a word association test. Those who generate only a few words in response to the stimulus have a *steep* associative hierarchy, whereas those who generate many have a *flat* associative hierarchy.

The relationship between defocused attention, sensitivity, and flat associative hierarchies makes sense. Recall that storage of memories is distributed across many memory locations, with degree of activation falling with distance from the most activated one. It seems sensible that the more stimulus features one attends, or is sensitive to, the more memory locations the current instant of experience gets stored to; that is, the more widely distributed the activation function. Therefore, the more memory locations from which ingredients for the *next* instant of experience can be drawn. Thus the greater the likelihood of catching a concept (or portion of one) that isn't usually associated with the experience that evoked it. So once the subject has run out of the more usual associations (e.g. CHAIR in response to TABLE), unusual ones (e.g. ELBOW in response to TABLE) come to mind.

Of course, flatter associative hierarchies are exactly what you would expect to result from a widely distributed activation function, and defocused attention may well be the means by which this achieved. Note that in a state of defocused attention or heightened sensitivity to detail, stimulus dimensions that are not directly relevant to the current goal get encodes in memory. Since more features of attended stimuli participate in the process of storing to and evoking from memory, more memory locations are activated, and these in turn activate more memory locations, *etc*. Therefore streams of abstract thought are more frequent, and last longer. So *if* a stimulus does manage to attract attention, it will tend to be more thoroughly assimilated into the matrix of associations that constitutes the worldview, and more time is taken to settle into, or collapse on, any particular interpretation of it. Thus, the less new stimuli can compete with what has been set in motion by previous stimuli, *i.e*. memory plays a larger role in conscious experience.

Another interesting consequence of defocused attention or heightened sensitivity is that a concept that lies in the periphery of the hypersphere of activated memory locations can *pull* the content of the next instant of experience quite far from the one that preceded and evoked it. The mutual information amongst not only concepts, but consecutive instants of a stream of thought, is lower. So the conceptual network is not only penetrated more deeply, but also traversed more quickly. In other words, there is an increased probability that one thought will lead in a short period of time to a seemingly unrelated thought.

Over the long haul, a tendency toward defocused attention will yield a memory with regions of finer granularity. Since more time is taken up with one's own creative worldview weavings, less is left over to imitate others. As a result, the individual settles on a more 'self-made' worldview, and will perceive gaps in human understanding that others do not see, and that no existing idea, concept, or artifact seems to fill. And since the self-made worldview contains not just the rule, but the reason behind it, it can adapt more easily.

However, one would expect there to be a trade-off between processing provocative stimuli *thoroughly,* and remaining alert to *new* stimuli. For example, say that Ann notices that Wanda's book is blue and torn, but Ben just notices that it is blue. Ann's mind will associate the book with memories of torn things, whereas Ben's will not. Ann's knowledge that the book was torn may turn out to be useful. For instance, she might be more alert, and ready to respond appropriately, to future indications of Wanda's sloppiness. She might say it's just a hunch--she may no longer recall ever having seen the torn book--but the fact that she once registered it as torn reconfigured her conceptual network in a way that made her more alert to the potential. On the other hand, Ann might be so busy pondering the torn state of the book that she fails to notice a truck headed her way! If you are still actively processing something that happened a minute ago, you might not be as attentive to what is happening now.

In sum then, an individual who is prone to defocused attention will tend to have flatter associative hierarchies; this would be the expected visible manifestation of a widely distributed activation function. The dynamical counterpart to (or consequences of) these traits is longer and more frequent streams of abstract thought, with lower correlation between one thought and the next. We will refer to this as heightened *conceptual fluidity*. The high conceptual fluidity individual will settle into a worldview which, although possibly lacking in some sorts of common knowledge, may be unique and richly-detailed in other respects.

## 5. THE CRYSTALLIZATION OF A CREATIVE IDEA

> Concrete is As concrete doesn't And voices can drown · Why should I Why should I Why should I Why should I · Solidify · Make me real · So¾ you¾ can¾see¾ me · I guess you thought I'd hide the sun from my liquid thoughts and Make ice for you
>
> --Cheryl Crow, from the song *Solidify*, on the album *Tuesday Night Music Club*.

We have looked at how an inkling surfaces to awareness through the defocusing of attention, and the consequent entry into a state of conceptual fluidity wherein unusual connections are gleaned. Now we examine the process by which this inkling solidifies into the form in which it will be received by the world.

### 5.1 Creations as Mirrors that Reflect and Reinforce the Self

I will begin with a proposal for what might motivate creative individuals to devote themselves to the process of shaping and refining their insights. Many will undoubtedly think this proposal stands on shaky ground. However, if the ground it stands on does eventually prove solid, it will help explain the urgency of the creative impulse. The ground is Chalmers' (1996) double aspect theory of information. According to this theory, whenever an information state in an information space is realized physically it is also realized phenomenally; that is, it has a conscious (or protoconscious) aspect. He uses the term information in the Bateson (1972) sense, as a difference that makes a difference to some interpretive (often intentional) system. Thus an informative event takes place when one thing interacts causally with another. The idea may sound implausible, but in fact it has much to recommend it, and it has been enormously influential (Shears 1997). So let us briefly explore the consequences of granting it provisional status.

The obvious question then is: what sort of architecture could coerce bits of phenomenally-endowed information to integrate their individual subjectivities into a single, concentrated subjectivity, such that the consciousness of non-self aspects of the world appear negligible by comparison? Biological systems are a perfect candidate because as autocatalytically closed systems, they are nonsymetrically inward-biased; they naturally trap and locally amplify information. Moreover, they generate not just novel informative components, but exactly those whose information-providing potential can be exploited by what is already in place, thereby enabling the system to function as a unified whole. We have seen how, in a human mind, where memories are integrated into a conceptual web through the emergence of abstractions, information may be further focused and integrated through a second tier of autocatalytic closure. This would not just amplify our own individual consciousness, it would tend to make us underestimate the degree to which non-self entities are conscious, where self is viewed in an extended sense; i.e., we tend to empathize with (experience what it feels to be) entities that are genetically (or memetically) similar to us (see Gabora 1999, in press, for details).

Now consider the case of individuals who are by nature so receptive that the phenomenal aspect inherent in incoming information constantly threatens to disrupt the stability of their worldview, including the pattern of information processing that generates their sense of self. In other words, their empathy is so exaggerated that their sense of self is readily absorbed into the external stimuli, and shifts every which way, depending on where they place their attention. Creative endeavors could ward off this kind of confusion, as follows. The sense of self, as expressed in painting, music, or whatever is being created, is reflected back at them as they watch their creation crystallize. This reinforces and strengthens the sense of self, like a beam of light hitting a mirror and reflecting back on and resonating with itself. By continually imbuing artistic materials with patterns that express the self, the creator counters the tendency of stimuli to overwhelm the self. Moreover, in the resulting reinforced state, the self is better able to resist being destabilized by the phenomenal aspect of external stimuli.

### 5.2 Variable Fluidity as the Crux of Creative Potential

What are the cognitive mechanisms underlying the refinement of a creative idea? There is a considerable body of research suggesting that creativity is associated with not just increased fluidity, nor just increased control, but both. As Feist (1999) puts it: "It is not unbridled psychoticism that is most strongly associated with creativity, but psychoticism tempered by high ego strength or ego control. Paradoxically, creative people appear to be simultaneously very labile and mutable and yet can be rather controlled and stable" (p. 288). As evidence in favour of this view, Feist cites (Barron 1963; Eysenck 1995; Fodor 1995; Richards et al. 1988; Russ 1993). He notes that, as Barron (1963) put it over 30 years ago: "The creative genius may be at once naïve and knowledgeable, being at home equally to primitive symbolism and rigorous logic. He is both more primitive and more cultured, more destructive and more constructive, occasionally crazier yet adamantly saner than the average person" (p. 224). There is also evidence of an association between creativity and variability in physiological measures of arousal. High creativity has been found to be correlated with variability in heart rate (Bowers and Keeling 1971), spontaneous galvanic skin response (Martindale 1977), and EEG alpha amplitude (Martindale and Hasenfus 1978). Thus creativity seems to involve the ability to vary the degree of conceptual fluidity in response to the demands of any given phase of the creative process.

### 5.3 Annealing on a New Worldview

The basic idea of varying the degrees of freedom as the creative or problem solving process proceeds is seen in simulated annealing, a mathematical technique for solving problems with multiple, frequently conflicting criteria. This kind of problem is particularly difficult to solve because the trade-offs between the multiple objectives or constraints are often unknown. The best that can be done is to find what is referred to as a Pareto optimal solution: one that cannot be improved with respect to any one objective without worsening some other objective (Steuer, 1986). A compromise is reached.

Simulated annealing was inspired by the physical process of annealing, during which a metal is heated and then gradually cooled. The shapes and charges of the atoms confer on them varying degrees of attraction and repulsion toward one another. Configurations that maximize attractive forces and minimize repulsive ones are most stable; thus, the shapes and charges of the atoms define the constraints of the system. Heating decreases the stability of the forces that bind the atoms together−−it loosens global structure−−whereas cooling has the opposite effect. The slower the cooling process, the more opportunity the atoms have to settle into a stable, low-energy arrangement. Simulated annealing computationally mimics the cooling process by decreasing and then gradually increasing the stability of the connections amongst the parts of a system in a series of either random or deterministic updates. If there are too few update steps, the system settles on a state wherein few of the constraints imposed by the structure and dynamics of its components are met. It may, for example, result in islands of mutually compatible components which are themselves incompatible, in which case the system has difficulty functioning as a whole. The greater the number of updates, the more harmonious the state the system eventually settles into, i.e., the more likely it is to find a Pareto optimal solution. An interesting feature of this process is that as the temperature is lowered, the correlation length−−that is, how far apart the components of a system must be before their mutual information falls to zero−−increases. The result is that a perturbation to any one component can percolate through the system and affect even distant components.

Simulated annealing has been used to improve the performance of a neural network (Cohen 1994; Cozzio 1995). The evidence of variable fluidity suggests that something analogous happens in the human mind. When inconsistencies abound, large-scale worldview renovation is in order. It seems reasonable to expect that at these times there is a tendency to temporarily loosen one's internal model of reality, weaken inter-concept relationships, so as to allow new insights to more readily percolate through and exert the needed revolutionary impact. Then one slowly "anneals" as the details of how to best structure this new and improved worldview fall into place. This could be achieved by decreasing the neuron activation threshold; that is, increasing the width of the distribution function, so that there is a higher degree of conceptual fluidity. Thus there is an increased potential for any thought to trigger a chain reaction of novel associations. The neuron activation threshold then slowly increases, thereby stabilizing associations that are consistent and fruitful.

Recall the trade-off between thoroughly processing provocative stimuli, and remaining alert to new stimuli. Neither is inherently *better*. However the thorough processing option affords greater possibility of making a creative contribution to the world. Once the overall framework of a unique idea has been painted in the broad strokes characteristic of the high-fluidity individual, failure to be understood may motivate a decrease in fluidity. This affords more delicate control over what gets evoked and refined. The details can then be fleshed out, grounding the idea more firmly in consensus reality, so that when it is born it is less vulnerable and more understandable. In sparse regions of conceptual space one might expect memories to be more widely distributed (to encourage the formation of abstractions there), and in dense regions, narrower (to permit finer distinctions). By adapting the degree of fluidity to the density of concepts in the region of the worldview being explored, one can remain at the edge of chaos at all times. In other words, this enables the degree of connectivity amongst concepts from one instant to the next to remain relatively constant.

The more variable the degree of fluidity, the more hierarchical levels at which inconsistency can manifest. Thus the more detailed (both fine/course-grained) the memory can potentially become. However, in general one can expect that the more hierarchical levels that must be climbed to achieve closure, the longer it takes for closure to be achieved. By analogy, a spider web can be both bigger *and* most tightly woven, but that takes longer. So it may take longer for the creative individual's worldview to become integrated such that there exists an association pathway from any one item in memory to any other. Nevertheless, when their memories finally *do* get woven into a unified whole, the resulting worldview is not only more unique and self-made, but it *also* captures more of richness and subtlety of the world we live in. Support for this comes from the finding that the less rule-driven, and thus the more contextual the domain, the older the age at which performance peaked; in other words, the longer it took for a unique closure to be achieved (Simonton 1984, 1999).

**5.4 Gestation and Birth: From Inkling to Insight**

A large part of the annealing process involves winnowing the insight out from the specifics of the memories or concepts originally involved in its conception, and making it applicable to the goals and modes of expression available for its manifestation. For example, the *essence* of the newly conceived TROPICALE construct is gleaned by reconsidering it in light of its various properties (such as fruity), non-properties (bitter, say), relationships (such as to your only direct experience of the tropics, a visit to the Bahamas), and instances (such as your imagined preconception of bottles of TROPICALE lined up on the grocery shelf), and then funneled through the constraints inherent to the domain of logo design. The degrees of freedom−−that is, the variety of directions the project can take−−gradually decrease. For example, as you zero in on the idea of drawing the T in TROPIC in the shape of a palm tree, alternative T fonts are ruled out.

When an idea has closed a gap at the level of the one's *own* worldview, it is ready to be nurtured to a state where it will close the corresponding gap at the cultural level. As the idea is refined, the creator's associative hierarchy can afford to narrow, and in fact *must* narrow, if is to be received by the world. Much as a beam of light diverges when it passes through a concave surface, a creative idea seems to lose clarity when expressed to others. So the existence of this second level of acceptance prompts the need for further crystallization. In the case of your Goosehead Breweries assignment, this involves progressively honing in on your most potent final draft of the TROPICALE logo, surrounding it with a flourish of fruit slices (perhaps, as a final added touch, making them vaguely genitalia-like so as to invite subliminal inklings in the brains of the buying public), and placing the finished product on your boss's desk.

## 6. CLOSING THOUGHTS

It may be that this chapter has barely touched on what lies at the crux creativity. It may turn out that religious and spiritual traditions have more to offer here of explanatory value than does science. Nevertheless, the notion that cultural, as well as biological information *evolves* through variation and convergence surely gives new vitality to our understanding of the creative process, both at the individual level and the cultural.

Of course, there are significant differences between biological and cultural novelty. In the biological realm, novelty is generated by combining and mutating information units. But in the cultural realm, all sensory information gets more or less thrown into one big melting pot (or keg, you might say), that is, the conceptual network, or worldview, and novelty is generated *strategically and contextually*, by highlighting those aspects of the worldview most relevant to the current situation. It may be that the origin of the capacity to weave memories into a worldview through the formation of abstractions gave rise to the origin of culture. And you could say that creativity is the crystallized precipitate of these worldview weavings.

For the recurring example used in this chapter, I did not chose Einstein or Mozart but an everyday individual, a graphic designer at a Brewery, to emphasize the point that *everyone* is creative. Every inkling in every mind alters, however minutely, the structure of a worldview which, through the spoken word, the holding of hands, *etc*., interacts with other worldviews, and is part of a vast chain of worldviews evolving through space and time. Nevertheless it is interesting to ask: was Einstein a seven-pack? Or was he a regular six pack like most of us, but one in which the plastic thingy melted and resolidified in a truly exceptional way?

Probably both. But the second is more likely what led him to be "person of the century" (according to Time magazine). The greater the number of concepts, the more levels of hierarchically structured abstraction are possible, but only *some* of these capture regularity of the world and therefore have meaning. So once there's a certain number of concepts in there, what matters most is how the individual weaves them together. In fact, up to an IQ of 120, intelligence and creativity are correlated, but past 120 they are not (Barron 1963). My guess is that Einstein had a tendency toward defocused attention and conceptual fluidity, and as a result, his worldview was less a product of what he was taught, more of a self-made construction from the bottom up. And the worldview weavings of the human race will, of course, never be the same.

## ACKNOWLEDGEMENTS

I would like to thank Beverlee Moore for comments on the manuscript, and acknowledge the support of the Center Leo Apostel and Flanders AWI-grant Caw96/54a.